\documentclass[12pt,preprint]{aastex}

\newcommand{\Teff}{\mbox{$T_{\rm eff}$}}
\newcommand{\logg}{\mbox{$\log g$}}

\newcommand\vmac{\mbox{$v_{\rm mac}$}}

\newcommand{\vsini}{\mbox{$v\sin i$}}
\newcommand{\kms}{\mbox{km s$^{-1}$}}

\slugcomment{DRAFT}

\begin{document}

\title{\bf Testing the Reality of Strong Magnetic Fields on T Tauri Stars:
           The Naked T Tauri Star Hubble 4}

\author{Christopher M. Johns--Krull\altaffilmark{1,2}}
\affil{Department of Physics \& Astronomy, Rice University, 6100 Main St.
       MS-108, Houston, TX 77005}
\email{cmj@rice.edu}

\author{Jeff A. Valenti\altaffilmark{1}}
\affil{Space Telescope Science Institute, 3700 San Martin Dr., Baltimore, MD
       21210}
\email{valenti@stsci.edu}

\author{Steven H. Saar\altaffilmark{1}}
\affil{Center for Astrophysics, Harvard University, 60 Cambridge Street,
       Cambridge, MA 02138}
\email{saar@head.cfa.harvard.edu}

\altaffiltext{1}{Visiting Astronomer, Infrared Telescope Facility, operated
for NASA by the University of Hawaii}
\altaffiltext{2}{Visiting Astronomer, McDonald Observatory, operated
by The University of Texas at Austin}

\begin{abstract} 

High resolution optical and infrared (IR) echelle spectra of the naked 
(diskless) T Tauri star Hubble 4 are presented.  The K band IR spectra include 4
Zeeman sensitive \ion{Ti}{1} lines along with several magnetically insensitive
CO lines.  Detailed spectrum synthesis combined with modern atmospheric
models is used to fit the optical spectra of Hubble 4 in order to determine
its key stellar parameters: \Teff = $4158 \pm 56$ K; \logg = $3.61 \pm 0.50$;
[M/H] = $-0.08 \pm 0.05$; \vsini = $14.6 \pm 1.7$ \kms .  These stellar
parameters are used to synthesize K band spectra to compare with the 
observations.  The magnetically sensitive \ion{Ti}{1} lines are all 
significantly broadened relative to the lines produced in the 
non-magnetic model, while
the magnetically insensitive CO lines are well matched by the basic
non-magnetic model.  Models with magnetic fields are synthesized and
fit to the \ion{Ti}{1} lines.  The best
fit models indicate a distribution of magnetic field strengths on the stellar
surface characterized by a mean magnetic field strength of $2.51 \pm 0.18$ 
kG.  The mean field
is a factor of 2.0 greater than the maximum field strength predicted by 
pressure equipartition arguments.  To confirm the reality of such strong 
fields, we attempt to refit the observed profiles using a two component 
magnetic model in which the field strength is confined to the equipartition 
value representing plage-like regions in one component, and the field is 
allowed to vary in a cooler component representing spots.  It is shown that
such a model is inconsistent with the optical spectrum of the TiO bandhead
at 7055 \AA.

\end{abstract}


\section{Introduction} 

Classical T Tauri stars (CTTSs) appear to be roughly solar mass pre-main
sequence stars still surrounded by disks of material which actively
accrete onto the central star.  Their properties have been recently
reviewed by M\'enard and Bertout (1999).  It is within the disks surrounding
CTTSs that solar systems form.  Understanding the processes through which
young stars interact with and eventually disperse their disks is critical for
understanding the formation of planets and the rotational evolution of stars.

     Over the past several years, strong stellar magnetic fields have come
to play a central role in current theories describing the interaction
of a CTTS with its circumstellar accretion 
disks.  Current theories posit that the stellar magnetic field
truncates the disk at or inside the corotation radius, directing
the flow of accreting disk material toward the polar regions of the
central star (e.g. Camenzind 1990, K\"onigl 1991, Cameron \& Campbell
1993, Shu et al. 1994, Paatz \& Camenzind 1996).  This magnetospheric
accretion model is driven by the need to explain how CTTSs can accrete
substantial amounts of material (and presumably angular momentum), yet
remain relatively slow rotators (Hartmann \& Stauffer 1989).  Early
observations showed that CTTSs later than K7 had rotation periods of 
7 - 10 days while diskless T Tauri stars (the naked TTSs: NTTSs) rotated
with a wide range of periods (Edwards et al. 1993).  While this dichotomy
in the rotation rates between CTTSs and NTTSs has recently been called into
question (Stassun et al. 1999) for stars in the Orion Nebula Cluster, it
is likely that some process must be regulating the rotation of CTTSs since
many of these stars do not appear to be spun up by disk accretion.  
The magnetospheric accretion model provides a means of regulating the stellar
rotation, and it has been used to successfully describe some
of the observed characteristics of high spectral resolution line profiles
observed in CTTSs (Hartmann, Hewett, \& Calvet 1994; Muzerolle, Calvet, \& 
Hartmann 1998).

     The most successful approach for measuring fields on late--type 
stars in general has been to measure Zeeman broadening of spectral lines in
unpolarized light (e.g., Robinson 1980; Saar 1988; Valenti, Marcy, \& Basri
1995; Johns--Krull \& Valenti 1996).  For any given Zeeman $\sigma$ 
component, the splitting resulting from the magnetic field is 
$$\Delta\lambda = {e \over 4\pi m_ec^2} \lambda^2 g B
                = 4.67 \times 10^{-7} \lambda^2 g B \,\,\,\,\,\, 
                {\rm m\AA\ kG^{-1}},\eqno(1)$$
where $g$ is the Land\'e-$g$ factor of the transition, $B$ is the
strength of the magnetic field (given in kG), and $\lambda$ is the
wavelength of the transition (specified in angstroms).
The first direct detection of a magnetic field on a TTS was
reported by Basri, Marcy, and Valenti (1992).  These authors used the
fact that marginally saturated magnetically sensitive lines will have
an increase in their equivalent width in the presence of a magnetic field
as the line components split to either side, increasing
line opacity in the wings.  Using this equivalent width technique, Basri
et al. (1992) measured the field of the NTTS Tap 35, obtaining a value of the
magnetic field strength, $B$, times the filling factor, $f$, of field regions
of $Bf = 1.0 \pm 0.5$ kG.  As the importance of magnetic fields on TTSs has 
become recognized, more effort has recently been put into their
measurement.  Guenther et al. (1999) used the same basic technique as 
Basri et al. (1992) to analyze spectra from 5 TTSs, claiming significant 
field detections on 2 stars.  

     From the perspective of cool star research, the high apparent field
strengths on TTSs is a little surprising.  Spruit \& Zweibel (1979)
computed flux tube equilibrium models, showing that magnetic field
strength is expected to scale with gas pressure in the surrounding
non-magnetic photosphere.  Field strengths set by pressure equipartition 
appears to be observed in G and K dwarfs (e.g. Saar 1990, 1994, 1996) and 
possibly in M dwarfs (e.g. Johns--Krull \& Valenti 1996).  TTSs have 
relatively low surface gravities and hence low photospheric gas pressures,
so that equipartition flux tubes would have relatively low 
magnetic field strengths compared to cool dwarfs.
Indeed, Safier (1999) examined in some detail the ability of TTS photospheres
to confine magnetic flux tubes via pressure balance with the surrounding 
quiet photosphere, concluding that the maximum field strength allowable on
TTSs is substantially below the current detections.  Safier does point out 
a number of potential problems in the data analysis which might affect the 
results of Basri et al. (1992) and Guenther et al. (1999) and lead to an 
overestimate of $Bf$ on these stars.  On the other hand, it is doubtful
that pressure equipartition arguments should hold when the magnetic field
filling factor approaches unity on the stellar surface.  Saar (1996)
finds that the magnetic field strength does rise above the pressure
equipartition value for very active K and M dwarfs once the filling factor
reaches a value near one.  As a result, super-equipartition fields on
TTSs, if confirmed, would be an extention of the behavior of very active
main sequence stars.

     Johns--Krull, Valenti, and Koresko (1999b - hereafter Paper I) have
examined three (K\"onigl 1991, Cameron \& Campbell 1993, Shu et al. 1994) of
the magnetospheric accretion flow theories which provide detailed analytic
expressions for the surface magnetic field strength on CTTSs.  In the
context of these theories, the surface field strength depends on the
stellar mass, radius, rotation rate, and the mass accretion rate onto
the star from the disk.  Taking values of these quantities from the
literature, Paper I shows that the predicted magnetic
field strengths vary over a large range of values, including strengths as
high as 11 kG.  The mass accretion rate estimates used in Paper I
are preferentially those of Hartigan, Edwards, and Ghandour
(1995).  Gullbring et al. (1998) argue that these accretion rates are
an order of magnitude too high.  Paper I showed that
the predicted field strengths depend on the square root of the accretion
rate.  Adopting the lower accretion rates favored by Gullbring et al.
(1998) suggests that the fields on CTTSs should range up to $\sim 4$ kG,
still quite a measurable value.

     Paper I made use of the wavelength dependence
of the Zeeman effect (see Eq. [1]) to detect actual Zeeman broadening of a
\ion{Ti}{1} line at 2.2 $\mu$m on the CTTS BP Tau, measuring an average 
field strength of $\bar B = 2.6 \pm 0.3$ kG.  The detection of Zeeman 
broadening yields magnetic field strengths largely immune to the 
uncertainties pointed out by Safier (1999) in equivalent width analyses; 
however, Paper I points out that the observations could be reproduced by
an accretion disk with unusual properties.  Paper I further suggests that
observations of additional IR lines (particularly magnetically insensitive
CO lines) are needed to validate the magnetic field detection.  
Paper I also points out that the mean field detected on BP Tau is larger 
than that which can be confined by pressure balance with the quiet photosphere
as discussed by Safier (1999); however, this is not a problem if the magnetic
filling factor is unity in the visible photosphere (see also Solanki 1994,
Saar 1996).  Indeed, the filling factor is unity in the dipole field 
configurations favored in current theoretical models (K\"onigl 1991, 
Cameron \& Campbell 1993, Shu et al. 1994).

     Another method for detecting magnetic fields is to look
for net circular polarization in Zeeman sensitive lines.  Observed along
a line parallel to the magnetic field, the Zeeman $\sigma$ components
are circularly polarized, with the components split to either side of
the nominal line wavelength having opposite helicity.
If there is a net longitudinal component of the magnetic field on the 
stellar surface, net circular polarization results, and is usually easier 
to detect than Zeeman broadening of optical lines in unpolarized light.
If the magnetic topology on the star is complex and small scale (as on the
Sun), nearly equal amounts of both field polarities will generally be present 
producing negligible net polarization.  Johns--Krull et al. (1999a) detect net 
polarization in the \ion{He}{1} $\lambda$5876 {\it emission} line on
the CTTS BP Tau, indicating a net longitudinal magnetic field of 
$+2.46 \pm 0.12$ kG in the line formation region.  This \ion{He}{1}
emission line is believed to form in the post-shock region where
disk material accretes onto the stellar surface (Hartmann, Hewett, \&
Calvet 1994; Edwards et al. 1994).  If the accretion shock forms at
or above the stellar surface, this observation provides additional
evidence that super-equipartition fields do exist on at least one CTTS.

In an effort to further test the reality of strong magnetic fields on TTSs,
we present here an analysis of the NTTS Hubble 4.  We detect Zeeman broadening
in four \ion{Ti}{1} lines in the K band.  Since NTTSs do not have close
circumstellar accretion disks, there can be no confusing the Zeeman
broadening signature in magnetically sensitive IR lines with kinematic
broadening due to formation in a disk.  To further verify the magnetic nature
of the detected broadening, 
several magnetically insensitive CO lines are also observed in the IR which
show no broadening beyond that expected from stellar rotation.

Hubble 4 is a K7 NTTS which shows strong, circularly polarized radio
continuum emission which suggests the presence of large scale,
ordered magnetic fields in this star (Skinner 1993).  In the recent
compilation of Brice\~no et al. (2002), they adopt \Teff = 4060 K and
$L_{\rm bol} = 2.7 L_\odot$ for Hubble 4, implying $R_* = 3.4R_\odot$.
Using the pre-main sequence evolutionary tracks of Palla and Stahler
(1999), these values of the effective temperature and luminosity imply 
$M_* = 0.8 M_\odot$.  Combining this mass with the above radius gives
an expected gravity of log$g = 3.28$.

We present below optical and IR high resolution spectroscopy that
implies strong magnetic fields are indeed present on the surface of the 
NTTS Hubble 4.  The observations and data reduction are described in \S\ 2.  
In \S\ 3 the analysis of the data is presented: \S\ 3.1 examines the optical
data used to determine the basic atmospheric parameters of Hubble 4; while
\S\ 3.2 presents the analysis of the IR data to measure the magnetic 
properties of this star.  A discussion of the results is given in \S\ 4.

\section{Observations and Data Reduction} 

     We obtained optical spectra with the Sandiford cross-dispersed, 
echelle spectrometer (McCarthy et al. 1993) mounted at the Cassegrain focus 
of the 2.1 m Otto Struve telescope at the McDonald Observatory. A Reticon 
$1200 \times 400$ CCD was used to record each spectrum covering the 
wavelength interval 5803 -- 7376 \AA. The spectrograph slit was approximately
1.07 arcsec wide on the sky, which projected to about 2.14 pixels on
the detector, resulting in a spectral resolving power of 
$R \equiv \lambda/\delta\lambda = 56,000$.  These spectra were reduced 
using an automated echelle reduction package written in IDL that is described 
by Hinkle et al. (2000). Data reduction included bias subtraction, 
flat-fielding by a 
normalized flat spectrum, scattered light subtraction, and optimal extraction
of the spectrum.  Wavelengths were determined by fitting a two-dimensional
polynomial to $n\lambda$ as function of pixel and order number, $n$, for 
several hundred thorium lines observed in an internal (part of the 
spectrograph assembly -- post telescope) lamp.  In addition to observations
of Hubble 4, spectra were also obtained of a rapidly rotating hot star
at similar airmass to correct for contamination by telluric absorption lines.
Table \ref{obs} summarizes the optical observations.

    The K-band spectra presented here were obtained at the NASA
Infrared Telescope Facility (IRTF) in December 1997.  Observations were made
with the CSHELL spectrometer (Tokunaga et al. 1990, Greene et al.\ 1993) using
a 0\farcs5 slit to achieve a spectral resolving power of $R \equiv
\lambda /\Delta\lambda \sim 37,300$, corresponding to a FWHM of $\sim
2.7$ pixels at the $256 \times 256$ InSb array detector.  A
continuously variable filter (CVF) isolated individual orders of the
echelle grating, and a 0.0057 \micron\ portion of the spectrum was recorded
in each of 3 settings.  The first two settings (2.2218 and 2.2291 $\mu$m)
each contain 2 magnetically sensitive \ion{Ti}{1} lines.  The third setting
(2.3125 $\mu$m) contains 9 strong, magnetically insensitive CO lines.
Each star was observed at two positions along the slit separated by 
10\arcsec.  The K band spectra were reduced using custom IDL software
described in Paper I.  The reduction includes removal
of the detector bias, dark current, and night sky emission.  Flat-fielding,
cosmic ray removal, and optimal extraction of the stellar spectrum are
also included.  Wavelength calibration for each setting is based on a 3rd
order polynomial fitted to $n\lambda$ for several lamp emission lines 
observed by changing the CVF while keeping the grating position fixed.
Dividing by the order number for each setting yields the actual
wavelength scale for the setting in question.  In each wavelength setting,
we observed Hubble 4 and a rapidly rotating star to 
serve as a telluric standard.  In addition, observations were obtained
in the two \ion{Ti}{1} settings of the K7 dwarf 61 Cyg B.  This inactive
star serves as a good null field reference against which we can check
details of the atomic line analysis.  The IR observations are also
summarized in Table \ref{obs}.

\section{Analysis}

     The wavelength dependence of the Zeeman effect makes the \ion{Ti}{1} 
lines in the K band excellent probes of magnetic fields in cool
stars (e.g. Saar \& Linsky 1985).  The goal is to measure the magnetic 
field on Hubble 4 by modeling
the profiles of the K band absorption lines.  Since other line broadening
mechanisms are also potentially important, it is necessary to first demonstrate
that the appearance of the spectrum in the absence of a magnetic field can
be accurately predicted.  Several steps are required to achieve this goal, and
the work here builds on the results of Paper I.  First, the optical spectrum 
of Hubble 4 is fit assuming no Zeeman broadening since the magnetic broadening
is weak compared to rotational and turbulent broadening at these wavelengths.
This was shown to be true for BP Tau in Paper I, and the \vsini\ of Hubble 4
is larger than that of BP Tau, making this even more true for Hubble 4 
(assuming a similar average field $Bf$).  The
regions of the optical spectrum used to fit atmospheric parameters have 
precise atomic line data based on fits to the solar spectrum (Paper I).  In 
addition, these regions have been further refined in Paper I by fitting the
optical spectrum of 61 Cyg B, a main sequence star with the same spectral
class (K7) as Hubble 4.  We reject from the analysis of TTSs some spectral
features that are poorly fit in the solar spectrum,
presumably due to inaccurate atomic parameters for
lines that are weak or absent in the solar spectrum.  Fitting the optical
spectrum of Hubble 4 yields the stellar parameters \Teff, \logg, [M/H], 
and \vsini.  There are various estimates of these quantities in the
literature (see \S 1), but to guarantee self--consistency, these nonmagnetic
quantities are redetermined with the same radiative transfer code
and model atmosphere used in the magnetic analysis.  Using the Hubble 4
model, the nonmagnetic infrared spectrum is predicted, and obvious
excess broadening in the observed \ion{Ti}{1} line profiles is noted while
no such excess broadening is seen in the magnetically insensitive CO lines. 
This excess width in the \ion{Ti}{1} lines is then modeled as Zeeman 
broadening, using a polarized radiative transfer code.  
Matching the observed profile then gives the strength and 
filling factor of magnetic regions on the surface of the star.

\subsection{Fit to the Optical Spectrum}

The details of the spectrum synthesis and fitting procedure are given
in Paper I.  Briefly, spectra for the optical
wavelength regions are calculated on a grid (in \Teff, \logg, and [M/H])
of model atmospheres.  We solve for the stellar parameters (\Teff, \logg, 
[M/H], and \vsini) using the nonlinear least squares technique of
Marquardt (see Bevington \& Robinson 1992).  Spectra between
model grid points are computed using three-dimensional linear (with
respect to the model parameters \Teff , \logg , and [M/H]) interpolation of
the resulting spectra.  We interpolate the logarithm of the line depth 
rather than residual intensity, because the former quantity varies more slowly
with changes in model parameters, yielding more accurate interpolated
values.  The model atmospheres are the ``next generation" (NextGen) atmospheres
of Allard \& Hauschildt (1995).  Rotational broadening in Hubble 4 is large
compared to the effects of macroturbulence.  This makes it difficult to
solve for \vmac\ in Hubble 4, so a fixed value of 2 \kms\ is adopted
from an extrapolation to K7 of class IV stars given in Gray (1992).
In previous work (Valenti et al.\ 1998), it was found that microturbulence
and macroturbulence were degenerate in M dwarfs, even with very high
quality spectra.  Therefore, for the low turbulent velocities considered 
here, microturbulence is neglected, allowing \vmac\ to be a proxy for all
turbulent broadening in the photosphere.  Spectra are synthesized in
3 optical regions, centered on 6151 \AA, 6498 \AA, and on the TiO
bandhead at 7055 \AA, which provides excellent temperature sensitivity.
As described above, within these regions, only certain spectral features 
are used to constrain the model parameters.  Following Paper I, we only
use features which are modeled well in both the solar spectrum and the
spectrum of 61 Cyg B, a K7V reference star.

The first line of Table \ref{h4tab} gives Hubble 4 model parameters
determined by fitting the optical spectrum.
Observed and synthetic spectra for
the three wavelength intervals are shown in Figure \ref{modelh41}.
The Marquardt technique used to find the best fitting stellar parameters
also gives formal uncertainties, but the actual uncertainties in derived
parameters are completely dominated by systematic errors.  To realistically
estimate uncertainties in the derived parameters, the observations of
Hubble 4 were reanalyzed with each of the 11 atomic line wavelength regions
(intervals shown in Figure \ref{modelh41} between 6140 -- 6500 \AA\ with 
contiguous bold pixels) successively removed from the list of constraints. 
This procedure yields 11 additional values for each model parameter.  The
standard deviation of each derived parameter value provides an estimate of
possible errors due to a bad fit in any one of the spectral segments.
These uncertainty estimates are given in the second line of Table
\ref{h4tab}.  The order containing the 7055 \AA\ band head is always 
included in this procedure because this bandhead is such a strong
temperature constraint.

     The spectrum of 61 Cyg B can also be used to obtain an estimate of
uncertainties in the optical analysis as was done for BP Tau in Paper I.
In that paper, similar resolution spectra of 61 Cyg B covering the wavelengths
used in this paper were presented and analyzed.  Those data are reused
here.  First, the observed spectrum of 61 Cyg B is artificially broadened to
$\vsini = 14.6$ \kms (the value recovered for Hubble 4), using the standard
rotational broadening convolution technique (Gray 1992) with limb darkening
coefficients determined from the atmospheric models used to fit 61 Cyg B 
in Paper I.  The value of \vmac\ was fixed at the 
value from the original fit and the remaining parameters (\Teff, \logg, [M/H],
and \vsini) were solved for.  This experiment was repeated with
\vsini\ fixed, instead of \vmac.  Finally, one last solution was determined
letting all parameters (including \vsini\ and \vmac) float.  These tests 
give some indication of how uncertainties in the parameters are affected
by moderately rapid rotation.  For each of the parameters solved 
for in Hubble 4, the uncertainty due to rotation is taken to be the maximum
difference in these parameters between the nominal fit to 61 Cyg B and
the fits obtained for the rotationally broadened spectrum.  These 
uncertainties are given in line 3 of Table \ref{h4tab}.  The total adopted
uncertainty is the sum of the uncertainties (in quadrature) produced 
by the scatter in the fits to the Hubble 4 data and that from the fits
to the rotationally broadened 61 Cyg B spectrum.  This is a
rather conservative estimator, but only a few of the possible sources 
of systematic error have been considered.  The final adopted
uncertainties for Hubble 4 are reported in the fourth line of Table
\ref{h4tab}.

\subsection{Fits to the Infrared Data}

The IR data considered in this paper fall into two distinct categories.  
The first consists of the CO lines at 2.3 $\mu$m.  
These lines are not magnetically sensitive.  In principle then, all
the parameters that determine their appearance (\Teff, \logg, [M/H],
\vsini, \vmac) are found from the optical fits.  These lines
serve as a check that the parameters determined in the 
optical accurately represent the IR spectrum of Hubble 4.  The second
category contains the magnetically sensitive \ion{Ti}{1} lines which
will allow a determination of the magnetic field on Hubble 4.  Each is
discussed in turn.

\subsubsection{The CO Lines}

\begin{figure}
\plotone{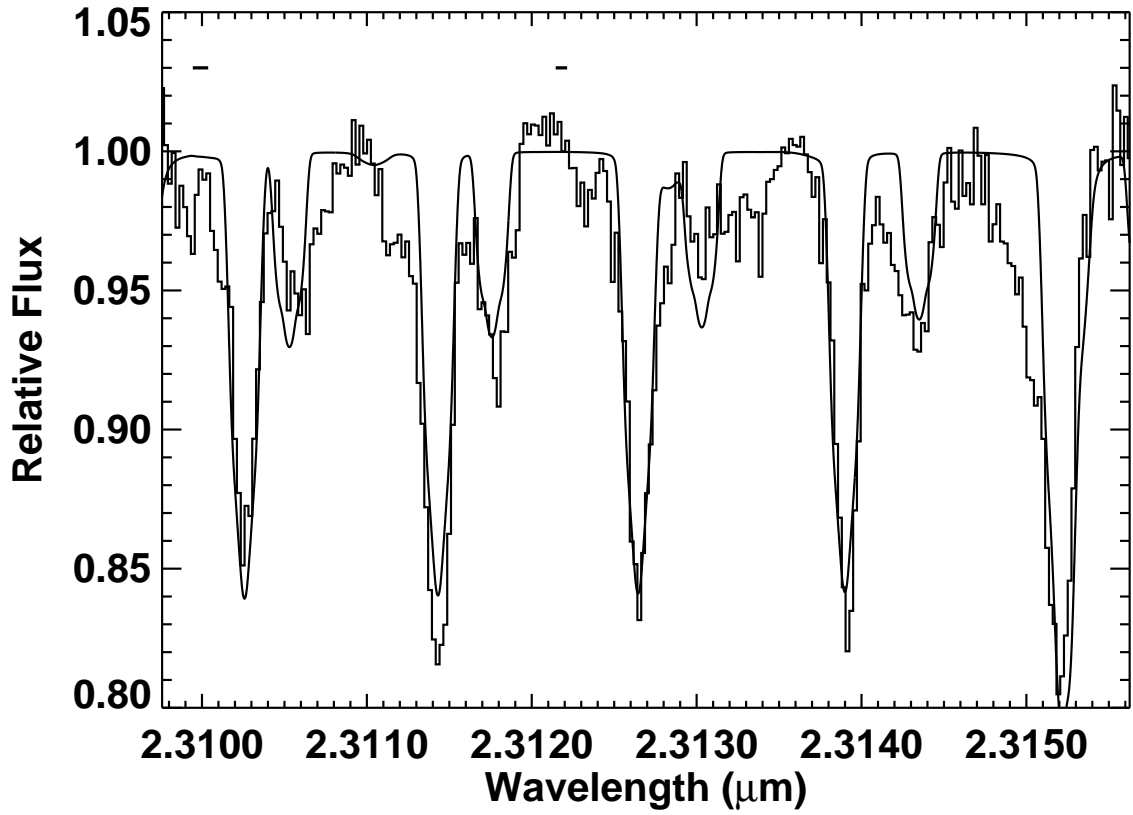}
\caption{\label{h4co}
The solid histogram shows the observed CO lines of Hubble 4.
The smooth solid curve shows model spectrum produced by fitting the
optical spectrum.  No fitting was done to the CO lines.}
\end{figure}

Line data for CO is taken from Goorvitch (1994) for the CO {\it X} 
$\Sigma^+$ state.  A model atmosphere is constructed by interpolating
the grid of NextGen model atmospheres
to the effective temperature, gravity, and metallicity determined above
for Hubble 4 (and reported in Table \ref{h4tab}).  The CO line spectrum
is computed using these parameters, and the comparison with the observed
CO lines is shown in Figure \ref{h4co}.  The agreement is excellent
within the signal-to-noise of the data.  In particular, the widths of
the CO lines is accurately predicted by the spectrum synthesis.  Again,
it is important to emphasize that no free parameters (aside from a 
continuum normalization value) are used to match the observed and
synthetic spectra.  The synthesis is determined entirely by the fits to
the optical spectrum of Hubble 4.  Examination of Figure \ref{h4co}
does show that the line strengths may be off slightly, no tuning of the
oscillator strengths has been attempted for the CO lines.  It is their
width that is of concern since these lines are insensitive to magnetic
fields, and the line widths are well fit by the model.

\subsubsection{The Magnetically Sensitive \ion{Ti}{1} Lines}

\subsubsubsection{\ion{Ti}{1} Oscillator Strengths}

Paper I only used one \ion{Ti}{1} line to model magnetic fields
in BP Tau.  BP Tau displays strong K band veiling (e.g.
Johns--Krull \& Valenti 2001), which made it impossible to distinguish
between the effects of veiling and inaccurate oscillator
strengths for the line.  In the case of Hubble 4, there is no K band
veiling and 4 \ion{Ti}{1} lines are available to model.  It is often the
case that inadequate precision in the laboratory data or theoretical
calculations results in oscillator strengths which produce synthetic lines
that are a poor match to the observed line depths.
As a result, some effort was expended to check the oscillator strengths for
all 4 lines as was done for the optical lines in Paper I.
Matching lines in the solar spectrum is a popular method to refine
oscillator strengths.  Unfortunately, the 4 \ion{Ti}{1} lines are very weak
in the solar spectrum, ranging in strength from $\sim 3.50 - 10.86$ m\AA\ 
based on the KPNO disk center solar atlas (Livingston \& Wallace 1991).  
The S/N of the disk center atlas becomes an issue when measuring such weak 
lines, particularly for the 2.2310 $\mu$m line which is the weakest of the 
four.  Given the weakness of these \ion{Ti}{1} lines in the solar spectrum,
the \ion{Ti}{1} $gf$ values were checked by comparison with the
spectrum of 61 Cyg B where the lines are much stronger.  In addition, using
61 Cyg B to constrain the oscillator strengths serves as a good self 
consistency check for the synthetic spectra of Hubble 4 since both stars 
have the same spectral type (K7), and 61 Cyg B is modeled using the same 
NextGen atmospheres used to 
model Hubble 4.  Atmospheric parameters for 61 Cyg B are taken from Paper I.
Initially, the synthetic \ion{Ti}{1} lines were weaker than the observed
lines in 61 Cyg B.  The $gf$ values of the 4 \ion{Ti}{1} lines were adjusted
and synthetic models were calculated and compared to the observed profiles
until $\chi^2$ was minimized.  The resulting $gf$ values are reported in
Table \ref{tigf} and are used to model Hubble 4.

\subsubsubsection{Single Field Component Model}

The simplest model to consider with a magnetic field is one in which some
fraction of the surface is covered by a field of a single strength,
while the rest of the surface is field free, or {\it quiet}.  Such a 
model is then specified by the field strength, $B$, and the filling
factor, $f$, of the field on the surface.  By analogy to the mean field
direction in solar plage, we assume a radial field geometry in the 
photosphere for Hubble 4.  On the Sun, strong
magnetic field is found in both solar plage regions which are slightly
hotter and brighter than the quiet atmosphere, and in sunspots which 
are significantly cooler than the quiet atmosphere.  However, the filling
factors of magnetic field reported for TTSs to date (Basri et al. 1992,
Guenther et al. 1999, Paper I) are very large, making it unclear that the
distinction seen in the thermal structure of field regions observed on the 
Sun is appropriate for TTSs.  Additionally, magnetospheric accretion theory
suggests there is strong magnetic field covering the entire surface of
TTSs, again making it unclear that the distinction observed on the
Sun is appropriate.  On the other hand, observations of photometric and
line profile variability on NTTSs supports the existence of brightness
inhomogeneties on their surface (e.g. Herbst et al. 1987; Bouvier \& Bertout
1989; Strassmeier, Welty, \& Rice 1994; Joncour, Bertout, \& Menard 1994)
Exploring to what extent the data demand that the
entire star is covered by strong field is one of the goals of the current
study, and this question is investigated further below.  To start with then,
we model the single field component using an atmosphere that matches the
overall effective temperature of the star, so that magnetic regions are
neither hot like plage nor cool like spots.

\begin{figure}
\plotone{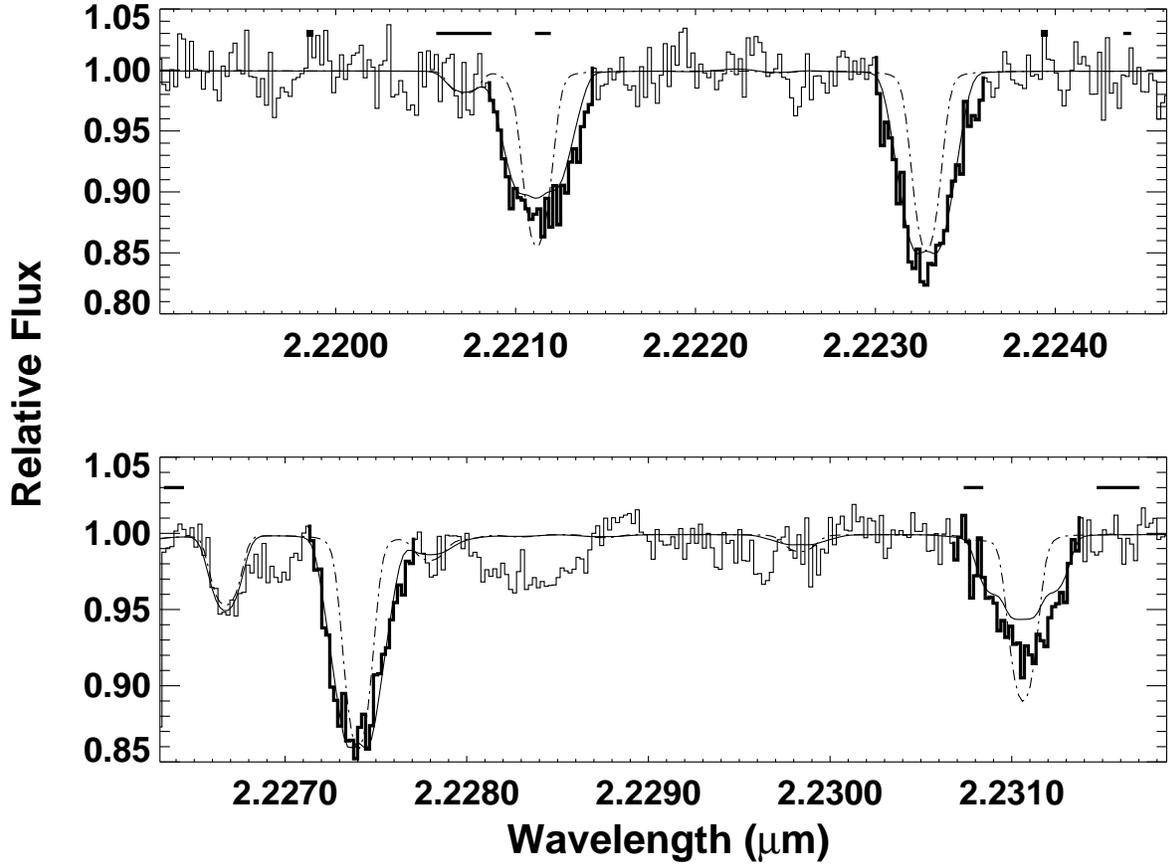}
\caption{\label{onecomp}
The solid histogram shows the observed \ion{Ti}{1} lines of Hubble 4.  The 
thick portions of the histogram show the regions of the spectrum actually
used in the fit.  The smooth solid curve shows the best fitting model
spectrum produced with a single magnetic region with $B = 2.98$ kG and
$f = 0.71$.  The dash-dot curve shows the spectrum produced by a
model with no magnetic fields.  Horizontal bold lines at relative flux of
1.03 indicate where telluric corrections exceeded 5\% of the relative flux.}
\end{figure}

The spectrum synthesis including magnetic fields has been described in
Paper I.  The Marquardt technique is again used to solve for the best 
fitting values of the field strength and filling factor.  The temperature,
gravity, metallicity, rotational and turbulent broadening parameters are 
held fixed at the values determined in the analysis of the optical spectrum.
The best fit to the IR data with a single field component model yields 
$B = 2.98$ kG and $f = 0.71$.  In Figure \ref{onecomp}, this model is compared
to the observed spectrum and a model with no magnetic field, and the fit 
parameters are also given in Table \ref{multi}.  Clearly, the 
model with a magnetic field is a much better fit to the observed spectrum,
yielding a reduced $\chi^2$ of $\chi^2_r = 1.95$ (computed 
considering only the points in the \ion{Ti}{1} lines shown in bold in Figure
\ref{onecomp}).  The non-magnetic model (which has no free parameters - 
everything is determined from the optical fits) gives $\chi^2_r = 12.68$.

\subsubsubsection{Multi-Component Field Models}

While this two component model provides a fairly good fit to the observed
lines, the computed 2.22112, 2.22328, and especially the 2.23106 $\mu$m lines
are all weaker than the observed profiles.  In Paper I, it was found that the 
best fit to the observed BP Tau \ion{Ti}{1} profile was achieved with a 
model consisting of a handful of separate magnetic components, each with 
a different filling factor.  This was motivated by the smooth, broad shape 
of the observed line profile which did not show resolved Zeeman $\sigma$ 
components.  In the case of Hubble 4 here, the higher $v$sin$i$ of this
star compared to BP Tau results in generally smooth profiles even when
only a single magnetic component is considered (Figure \ref{onecomp}).
However, one of the hallmarks of magnetic fields is that they increase
the equivalent width of partially saturated spectral lines which are Zeeman
sensitive (Leroy 1962, Basri et al. 1992), and this is true of the models 
shown in Figure \ref{onecomp}.  The exact increase in the line equivalent
width is a function of the line strength and the number and magnetic 
sensitivity of Zeeman components which split in response to the magnetic
field (Basri et al. 1992, Guenther et al. 1999).  This effect can be enhanced
even further if multiple magnetic components are considered.  Finally,
if the fields on TTSs are dipolar in nature as assumed by theory, a 
distribution of field strengths is {\it predicted} with the maximum field 
strength twice that of the minimum field value.  While observations do not
support a dipolar geometry for the fields (e.g. Johns--Krull et al. 1999b),
it is still interesting to consider multi-component magnetic field models 
for Hubble 4.

In the context of such models, it was found in Paper I that multicomponent
models with too many field components are degenerate since
the combined effects of the spectrometer resolution and stellar rotation
make it difficult to constrain separate field components if their field
strengths are too similar.  This effective magnetic resolution was found to
be $\sim 2$ kG in the case of BP Tau.  The larger \vsini\ of Hubble 4 acts 
to lower the effective magnetic resolution achievable on this star.  An
estimate can be calculated by adding in quadrature the broadening width, 
$\Delta\lambda$, due to the spectrometer resolution ($R = 37,300$) to the
broadening produced by rotation ($\Delta\lambda = v{\rm sin}i * \lambda/c$).
The resulting characteristic line width is $\Delta\lambda = 1.24$ \AA\ at
the wavelength of the K band \ion{Ti}{1} lines.  Putting this into equation (1)
and assuming a typical Land\'e-$g$ value of 2.0 for the Ti lines gives
a magnetic field strength of 2.68 kG.  This should be roughly the 
magnetic resolution achievable with the observations analyzed here;
however, simultaneous modeling of all 4 available \ion{Ti}{1} lines 
(with Land\'e-$g$ extending up to 2.50) 
mediates this effect somewhat.  As a result, our analysis of Hubble 4 below
has approximately the same effective magnetic resolution (2 kG) as our
previous analysis of BP Tau (Paper I).

The multi-component field models are constructed by calculating 
synthetic spectra at specific field strengths,
assigning a filling factor to each field component subject to the
constraint that the total filling factor is 1.0, and then summing up
the individual spectra weighted by their filling factors.  To fit the
observations the Marquardt technique is again used to find the
optimal combination of filling factors which produces the best fit to
the observed line profiles.  Since there is potential
degeneracy in the combination of filling factors, a Mont\'e Carlo technique
is used to test how sensitive the solution is to the assumed starting
value for each filling factor.  For each combination of filling factors
that are being solved for, the fit is performed 100 times with randomly
chosen starting values for the filling factors.  The mean and standard
deviation of each solved filling factor is then computed from these 100
different trials.  

A total of 4 different multi-component models are considered.  The model
results are summarized in Table \ref{multi}.  Model M1
solves for the filling factors of component fields of 0 -- 7 kG in
1 kG steps for a total of 8 different components.  The filling factors
and their standard deviations (for all models) are reported in Table 
\ref{multi}.  Also reported in the Table is the value of $\chi^2_r$ 
produced by the distribution of mean filling factors.  In all 4 models,
this value of $\chi^2_r$ is within 0.02 of the mean value of $\chi^2_r$ 
derived from the Mont\'e Carlo trials for a given model.  Lastly, the sum of 
the field strength multiplied by its filling factor ($\Sigma Bf$) is also
given.  This value corresponds to the mean field strength, $\bar B$, on the
stellar surface.  

Examination of model M1 shows substantial filling factors are found every
2 kG, with very little filling factor at the field values in between.  It
is unlikely that the true distribution on the star is so punctuated.  This
behavior is most likely a manifestation of the limited magnetic field
resolution discussed above.  Due to this limited magnetic resolution and
the behavior shown in model M1, model M2 was limited to only considering
field values from 0 -- 6 kG in 2 kG steps.  The results for model M2
are given in the third column of Table \ref{multi}.  

\begin{figure}
\plotone{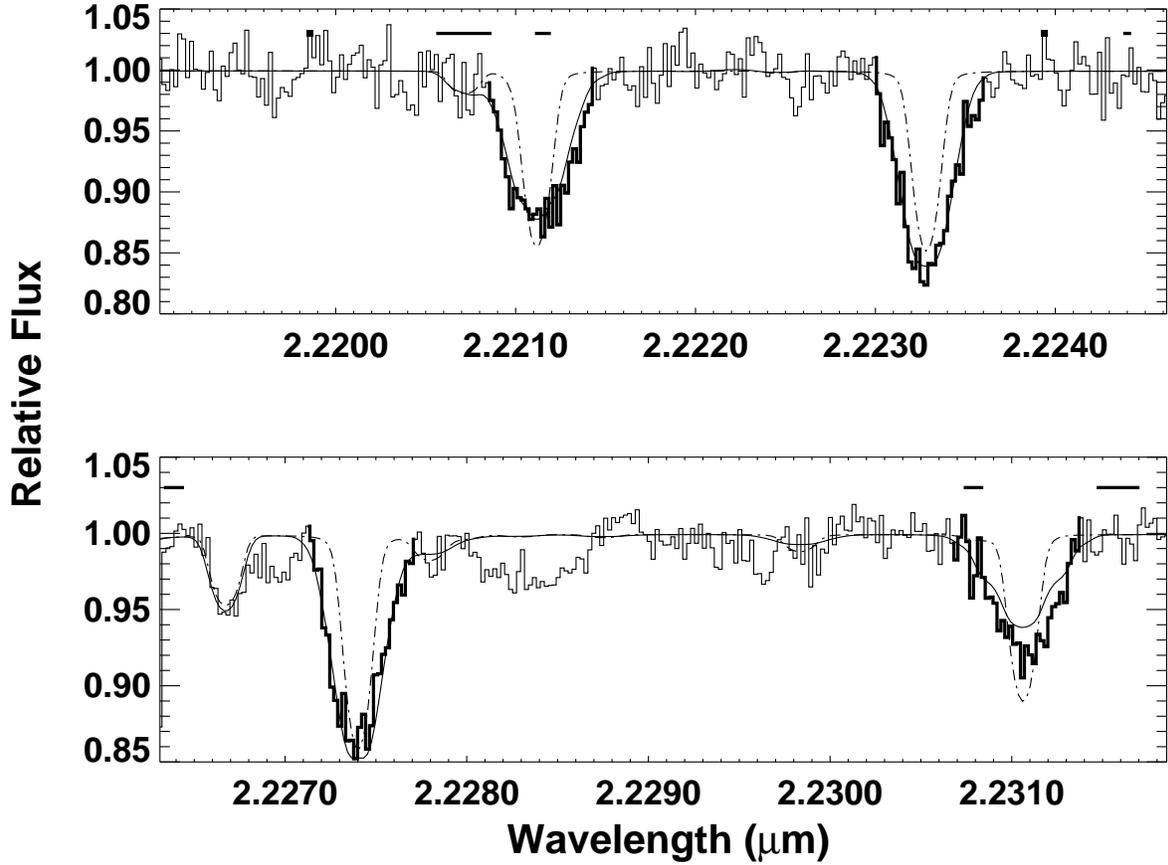}
\caption{\label{h4M4}
The solid histogram again shows the observed \ion{Ti}{1} lines of Hubble 4.
The thick portions of the histogram show the regions of the spectrum actually
used in the fit.  The smooth solid curve shows the best fitting spectrum
from the multi component model M4 of Table \ref{multi}.  Again,
the dash-dot curve shows the spectrum produced by a model with no magnetic
fields. Again, horizontal bold lines at relative flux of
1.03 indicate where telluric corrections exceeded 5\% of the relative flux.}
\end{figure}

An interesting question to consider is the extent to which a quiet ($B=0$)
field component is required by the \ion{Ti}{1} data.  As mentioned above,
current theories of magnetospheric accretion suggest the entire surface
of TTSs is covered by kilogauss level magnetic fields.  Model M1 is
indeed found to have only 1\% coverage on the stellar surface of field
free regions.  Therefore, model M3 was constructed only considering fields
in the 1 -- 7 kG range, again in 2 kG steps.  The distribution of filling
factors is reported in the fourth column of Table \ref{multi}.  While
somewhat marginal, the reduced $\chi^2$ of model M3 is indeed lower
than that for models M1 and M2.  Finally, examination of the results for
models M1, M2, and M3 reveals relatively little filling factor in regions
with field strengths above 5 kG.  Model M4 was therefore constructed 
to only consider fields in the range 1 -- 5 kG in 2 kG steps, and the
results are reported in the final column of Table \ref{multi}.  This 
model produced the lowest value for $\chi^2_r$ and is shown in
Figure \ref{h4M4}.

The Mont\'e Carlo trials for the multi-component models produce fits with 
nearly identical values of $\chi^2_r$ for a given model.  This indicates that
the $\chi^2_r$ surface is relatively smooth, and the results are essentially
insensitive to the initial guess for each model.  We also note that the trial
giving the minimum value of $\chi^2_r$ for each model gives filling factors
within one sigma of the mean reported in Table \ref{multi}.  These Mont\'e
Carlo trials do not represent the uncertainty in the actual magnetic flux,
however.  The difference in derived mean fields ($\Sigma B f$) for the
different models suggests that systematic uncertainties arising from the
model assumptions dominate the overall uncertainty.  We assign a value of
2.51 kG for the mean field, which is the value for the model with the
lowest $\chi^2_r$.  To conservatively estimate the uncertainty in this mean 
field, the standard deviation of the mean field resulting from all the models
of Table \ref{multi} except the spot field model (see below) is computed.  
This gives an uncertainty estimate of 0.18 kG for the mean field of Hubble 4.

\subsubsubsection{Spot Fields}

As described above, due to their lower surface 
gravity, the gas pressure in TTS photospheres is unable to confine magnetic 
fields stronger than about 1 kG, depending on the exact values of \Teff\ 
and \logg\ in the stellar atmosphere.  In addition, Safier (1999) pointed
out that methods based on line equivalent widths are very sensitive to 
errors in the assumed \Teff\ which might lead to errors in the estimated
field strength.  On the other hand, the field detected on BP Tau in Paper I
and on Hubble 4 above is measured by the excess {\it width} of magnetically
sensitive photospheric lines, and the field strength is insensitive to 
small errors in \Teff; however, it is still possible that the strong fields
are confined to starspots where the field can be larger due to the larger
depth of formation of the spectrum in a spot, if the solar analogy is assumed
to apply for starspots.  In this case though, the lower radiative surface
flux emerging from a cool spot compared to the quiet photosphere may require 
a corresponding increase in the filling factor of magnetic regions to produce
Zeeman broadened line wings of similar depth to those shown in Figure 
\ref{onecomp}.  This effect is offset by the fact that the \ion{Ti}{1}
lines in the K band are quite temperature sensitive, and will be substantially
stronger in a cooler atmosphere.  Thus, it is an open question whether the
strong fields observed on TTSs can be confined to cool starspots, with the
rest of the photosphere being field free or similar to plage.

\begin{figure}
\plotone{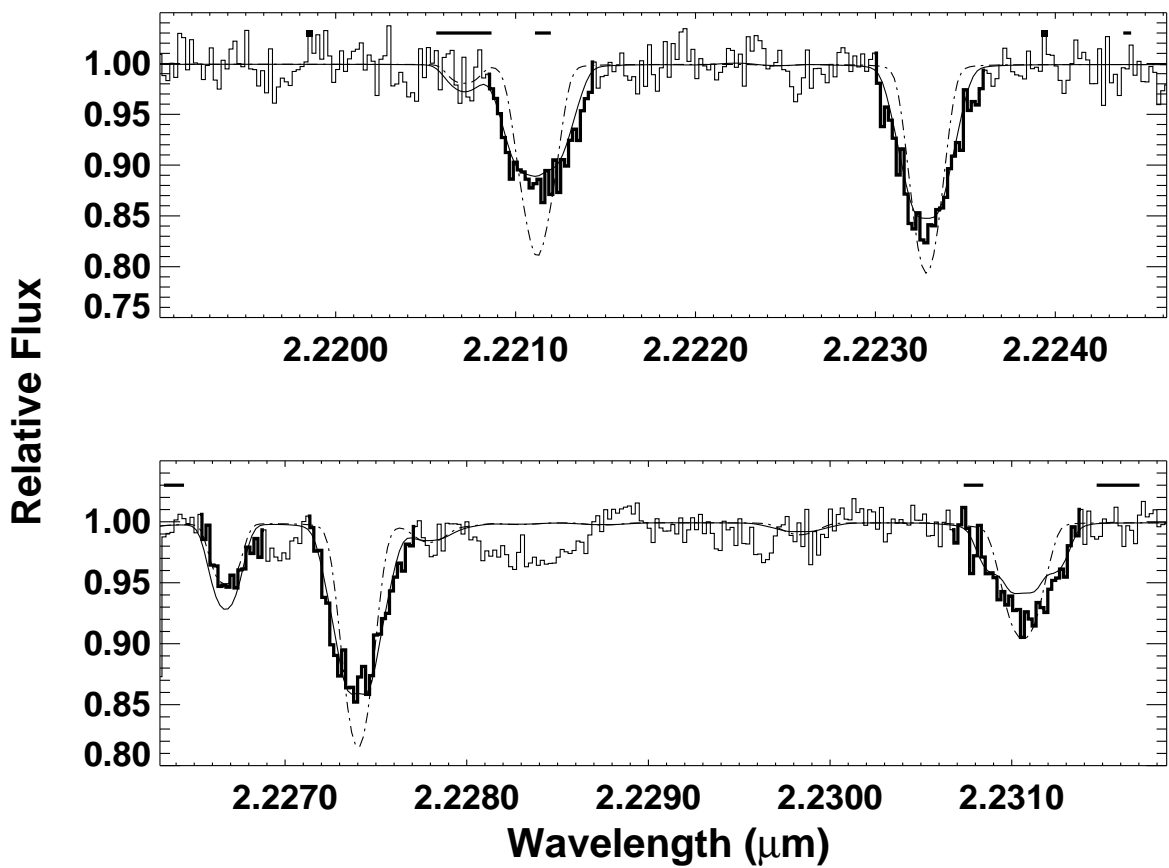}
\caption{\label{twocomp}
The solid histogram again shows the observed \ion{Ti}{1} lines of Hubble 4.
The thick portions of the histogram show the regions of the spectrum actually
used in the fit, now including the \ion{Sc}{1} line at 2.2266 $\mu$m 
which is used in the comparison of the spot model (see text).  The smooth 
solid curve shows the best fitting spectrum produced when the strong
magnetic fields are restricted to cool starspots.  The dash-dot curve shows
the spectrum produced the best fitting model in which the strength of
the magnetic field is restricted to 1.26 kG which is the value expected
when the fields are in pressure equilibrium with the surrounding quiet 
photosphere.  Again, horizontal bold lines at relative flux of
1.03 indicate where telluric corrections exceeded 5\% of the relative flux.}
\end{figure}

Following the solar analogy then, it can be assumed that a TTS atmosphere 
contains three components: field free regions, cool spots containing
strong fields, and plage-like regions containing field whose strength
is set by pressure balance with the surrounding quiet photosphere.
Solar plage shows up as brighter regions in strong lines such as H$\alpha$
and \ion{Ca}{2} H and K; however, these regions are generally not distinct
in the continuum and in weak atomic lines.  Therefore, it is assumed that
in the region of the K band \ion{Ti}{1} lines studied here, plage-like
regions on Hubble 4 have the same temperature structure as quiet regions
of the atmosphere.  The field in these regions is set by the gas pressure
in the quiet atmosphere.  This can be estimated using the Sun and setting
$B = B_\odot (P/P_\odot)^{1/2}$, where $P$ is the total gas pressure in 
the atmosphere.  In cool stars, $P \propto g^{2/3}$ where $g$ is the gravity
(Gray 1992).  Using \logg$ = 4.44$ for the Sun,
\logg$ = 3.61$ for Hubble 4 (from Table \ref{h4tab}), and taking 
$B_\odot = 1.70$ kG which is the maximum field reported by R\"uedi et al. 
(1992) for solar plage, the maximum plage field for Hubble 4 is 1.24 kG.  
Another estimate for this value can be made by setting the magnetic 
pressure equal to the gas pressure in the quiet atmosphere where the 
temperature is equal to \Teff.  Doing this, the plage field on Hubble 4 is 
estimated to be 1.26 kG.  These two estimates are quite close, and the higher
one (1.26 kG) is adopted for the analysis shown below.  The value is sometimes
referred to as the (pressure) equipartition field strength.  In the interest of 
exploring whether such a field strength is adequate to match the observations
of Hubble 4, an initial single field component fit was made fixing the magnetic 
field strength at $B = 1.26$ kG, allowing only the filling factor to vary.  
The best fit yielded $f = 0.92$, with $\chi^2_r = 8.02$.  This fit
is shown in Figure \ref{twocomp}.  Allowing field only equal to the 
equipartition value is not adequate to match the observations of Hubble 4.

The next step is to add a starspot component to the equipartition (plage-like)
component.  The spot field strength is treated as a free parameter; however,
it is necessary to fix the temperature of the spots relative to 
the quiet photosphere to a specific value.  Multi-color photometric monitoring
(see Herbst et al. 1994 and references therein) and Doppler imaging analysis
of TTSs (Strassmeier, Welty, \& Rice 1994; Joncour, Bertout, \& Menard 1994;
Joncour, Bertout, \& Bouvier 1994; Hatzes 1995; Rice \& Strassmeier 1996; 
Johns--Krull \& Hatzes 1997) finds a typical spot temperature $\sim 1000$ K
cooler than the quiet regions, so this temperature difference is assumed here.  

If it is assumed that all the field found above in the single field
component fit is in a cool spot
($f_s = 0.59$) with a temperature 1000 K below that in the non-magnetic
atmosphere, the temperature in the non-magnetic atmosphere must be
larger than 4158 K (\Teff\ found above) so that the overall effective
temperature is 4158 K.  If $T_q$ is the temperature in the quiet atmosphere
and $f_s$ is the filling factor of starspots, then this temperature is 
related to the stellar effective temperature by 
$(1 - f_s) T_q^4 + f_s (T_q - 1000)^4 = \Teff^4$ if the spot temperature
is fixed 1000 K below $T_q$, and the filling factors are area filling
factors (as they are in the synthetic modeling below).  A first guess is
obtained by setting
\Teff$ = 4158$ K and $f = 0.59$, which gives $T_q = 4660$ K.  Guenther
et al. (1999) point out that the filling factor of magnetic regions should
be higher if these regions are in fact cool spots due to the lower 
continuum flux.  This effect is minimized in the IR region where the
brightness contrast between a \Teff = 4660 K and \Teff = 3660 K atmosphere
is smaller than in the optical.  This effect is also counteracted for
the \ion{Ti}{1} lines studied here since they grow in strength substantially
as the temperature falls.  Therefore, a first attempt was made assuming a
spot temperature of $T_s = 3660$ K and a quiet and plage temperature of
$T_q = T_p = 4660$ K.  Again, the plage field strength is held fixed at 1.26
kG.  The spot field strength, $B_s$, and filling factor, $f_s$, as well as
the plage filling factor, $f_p$, are allowed to vary, and the quiet filling
factor is given by $f_q = 1 - f_s - f_p$.  Synthetic models were computed on
a grid of NextGen atmospheres and these were interpolated
in the same fashion as done for the optical spectra (see Paper I) to
obtain spectra for the spot, plage, and quiet regions.  The Marquardt method
was used again to fit the \ion{Ti}{1} line profiles.  The spots and plage-like
regions are assumed to be evenly distributed over the stellar surface.
The best fit is shown in Figure \ref{twocomp} and has $f_q = 0.42$, 
$f_p = 0.00$, $f_s = 0.58$, and $B_s = 2.97$ kG.  The
average field on the stellar surface is then $\Sigma Bf = 1.72$ kG.  This 
fit gives $\chi^2_r = 1.73$, equal to the fit of model M4 in Table 
\ref{multi}.  The fit parameters are also given in Table \ref{multi}.

The quality of the spot fit suggests that the strengthening of
the \ion{Ti}{1} line in the cool spot atmosphere completely offsets the
lower continuum brightness of the spot region relative to the quiet
region.  It is surprising that the best fit is obtained with only spot
and quiet regions, without any plage like regions on the surface.  At the
relatively low field (1.26 kG) for the plage like regions, the dominant
effect on the spectrum is to enhance the equivalent width of the lines
somewhat.  Due to the large \vsini\ of Hubble 4, this makes the \ion{Ti}{1}
lines slightly deeper, but not appreciably broader.  However, the Ti
lines grow significantly in strength in the relatively cool atmosphere of
the spot regions.  Therefore, there is no need to include plage like regions
to obtain the best fit.  Recall also, that the multi-component fits
indicate our magnetic resolution is $\sim 2$ kG, so it appears the plage
like component is simply degenerate and is consistently driven to zero
filling factor by the fitting routine.

In addition to the \ion{Ti}{1} lines visible in Figure \ref{twocomp}, 
attention should be paid to the \ion{Sc}{1} 2.2266 $\mu$m line.  In
the fit allowing for cool spots, this \ion{Sc}{1} line is noticeably
stronger than in the observations and is stronger than in the single field
component and multi component models.  Many \ion{Sc}{1} lines in this
region of the spectrum possess a significant nuclear spin which results
in hyperfine splitting of the lines.  We inspected a high resolution
($R = 225,000$) FTS spectrum of this line retrieved from the NSO
Digital Library (http://diglib.nso.edu), and find that this line displays
very small hyperfine structure.  In addition, this line is relatively weak 
in Hubble 4, so hyperfine splitting has only a small effect on the
equivalent width (though we note, any increase in equivalent width would
occur in both the M4 and spot models, with the spot model still a poorer 
fit).  All together, the influence of hyperfine splitting has less than
about a $\sim 5$\% effect on the line equivalent width.  Such splitting
is ignored here, and the line is treated as magnetic following LS 
coupling.  The poorer fit of the spot model relative to all the non-spot
models in the \ion{Sc}{1} line
suggests that the reality of such large spot regions can be tested by
strongly temperature sensitive lines.  The $\chi^2_r$ values quoted above
for all the magnetic models are based only on pixels encompassing the 
\ion{Ti}{1} lines.  If the \ion{Sc}{1} line is also included, the spot model
above gives $\chi^2_r = 2.19$ (\ion{Sc}{1} pixels used for $\chi^2_r$ 
calculation are shown in Figure \ref{twocomp}).  On the other hand,
model M4 of Table \ref{multi} and Figure \ref{h4M4} gives $\chi^2_r = 1.94$
when including the \ion{Sc}{1} line.  This suggests the spot model
is not the best description of the fields on Hubble 4.

\begin{figure}
\plotone{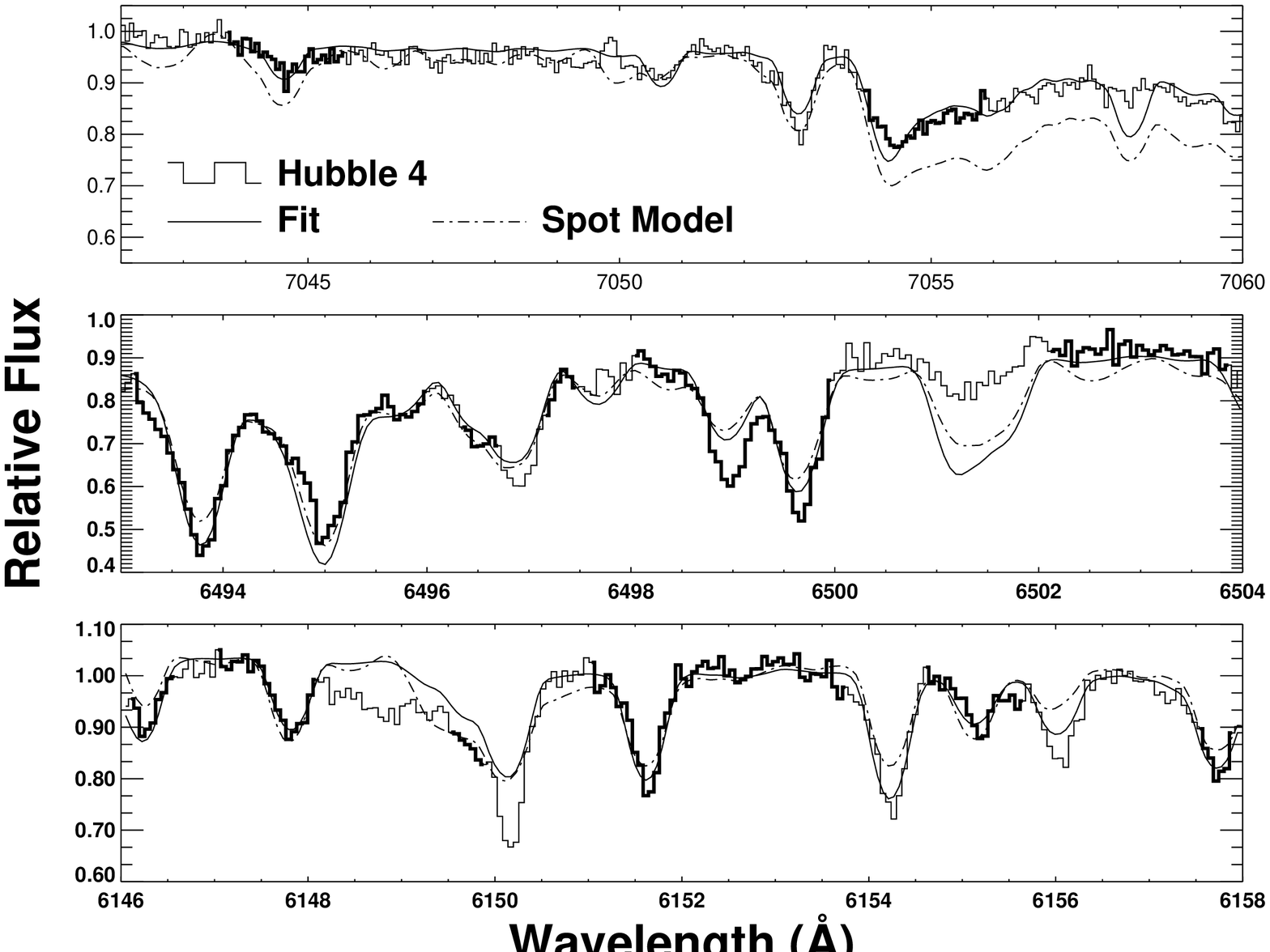}
\caption{\label{modelh41}
The solid histogram shows the observed optical spectrum of Hubble 4.  The
thick portions of the histogram show the regions of the spectrum actually
used in the fit.  The smooth solid curve shows the best fitting model
spectrum.  The dash-dot curve shows the spectrum produced by the
model in which strong fields are confined to cool starspots.}
\end{figure}

A more sensitive test of the spot model just described is to look again
at the optical spectrum analyzed in \S 3.1.  The TiO bandhead at 7055 \AA\
is a very temperature sensitive diagnostic at a spectral type of K7.
The optical spectrum for the spot model shown in Figure \ref{twocomp}
is shown in Figure \ref{modelh41}.  This model was computed without
magnetic fields, and thus represents only the effect of the two 
temperature components.   In Paper I, it was shown that strong magnetic
fields such as those found above for Hubble 4 have a very small effect on
the appearance on the optical spectrum in these regions for stars with
large $v$sin$i$ ($> 10$ km s$^{-1}$) as the case here.  In addition, the
original fits shown in Figure \ref{modelh41} are also calculated 
ignoring the magnetic field.  As quoted in \S 3.1, the original fits
produce a reduced $\chi^2 = 5.20$, whereas the spot model fit of 
Figure \ref{modelh41} give reduced $\chi^2_r = 12.62$.  Clearly, such
a spot model is not a better fit to the overall spectrum.  

\section{Discussion}

     Using high resolution optical spectra, we have determined atmospheric
parameters (Table \ref{h4tab}) for Hubble 4 which agree well with most
published estimates.  As
described in \S 1, Brice\~no et al. (2002) adopt $T_{\rm eff} = 4060$ K
and $L_{\rm bol} = 2.7 L_\odot$ for Hubble 4, implying $R_* = 3.4R_\odot$.
Using the pre-main sequence evolutionary tracks of Palla and Stahler
(1999), these values of the effective temperature and luminosity imply
$M_* = 0.8 M_\odot$.  Combining this mass with the above radius gives
an expected gravity of log$g = 3.28$.  The fit to the optical spectrum
finds $T_{\rm eff} = 4158 \pm 56$ K and log$g = 3.61 \pm 0.50$ in good
agreement with these numbers.  The spectroscopically determined 
temperature can be combined with the luminosity above (assuming a 10\%
uncertainty) to find $R_* = 3.19 \pm 0.33$.  This in turn can be used
with our spectroscopic estimate of gravity to obtain $M_* = 1.5 M_\odot$
with a factor of
1.8 uncertainty, driven almost entirely by the uncertainty in the 
fitted gravity.  While not a precise measure of the stellar mass, it 
is interesting to note that the mass determined here is larger than
that estimated from evolutionary tracks.  The same result was found for
BP Tau in Paper I, albeit with similarly large uncertainty.  In their
study of stars with dynamical mass determinations, Hillenbrand and White
(2004) also find that evolutionary tracks tend to underpredict the 
mass in lower mass stars.

In Paper I, clear excess broadening of the 2.2233 \micron\ \ion{Ti}{1} line
was interpreted as Zeeman broadening due to the presence of a mean
magnetic field of 2.6 kG in the photosphere of BP Tau.  However,
that paper pointed out that a disk with unusual properties could in
principle produce the same kind line broadening as observed.  Here,
the NTTS Hubble 4 is studied which lacks a close circumstellar disk.
Therefore, the near IR \ion{Ti}{1} lines from this star must be 
photospheric in origin.  As for BP Tau in Paper I, clear excess broadening
is seen in the 4 observed magnetically sensitive 2.2 \micron\ \ion{Ti}{1} 
lines relative to what is expected from a model calculation based on fitting
lines observed in the optical at high spectral resolution.  Each of the
four Ti lines has a different Zeeman splitting pattern and hence a
different predicted response to magnetic fields.  The observed line widths
and strengths respond as predicted for Zeeman broadening, adding support
for this interpretation.  The fit based on the optical spectrum accurately
reproduces magnetically insensitive 2.3 \micron\ CO lines.  The widths of
infrared CO lines and all optical lines are dominated by rotational 
broadening.  The infrared atomic lines show significant additional 
broadening that must be due to a process that preferentially affects the
IR lines.  Zeeman broadening is such a process.  Therefore, Zeeman broadening
due to strong stellar magnetic fields appears to be the only explanation
for the large linewidths of the \ion{Ti}{1} lines.

Modeling the observed Ti lines with magnetic fields produces good fits
to the observed spectra.  Unlike BP Tau in Paper I, the evidence demanding
the presence of multiple magnetic components with different field strengths
on the surface of Hubble 4 is not as strong, though it is present.  This 
appears to be the result of the larger \vsini\ of Hubble 4 (14.6 \kms)
compared to that of BP Tau (10.2 \kms), which yields somewhat poorer 
Zeeman resolution in our analysis of Hubble 4.  The best fitting single field
component field model gives $\chi^2_r = 1.95$, while the best fitting 
multi-component model (M4 of Table \ref{multi}) gives $\chi^2_r = 1.73$.
It is interesting to note that the best fit multi-component model
does not have a field free component.  Thus, with the current data it is
not possible to show that Hubble 4 possesses any field free regions.

The mean magnetic field on the surface is computed from the models by
summing the quantity $Bf$ over the different components used in the model.
This mean field we find for Hubble 4 is $2.51 \pm 0.18$ kG.
This field strength is comparable to the field 
strength required for some current magnetospheric accretion models (Paper I).
Equipartition arguments suggest that the maximum magnetic field strength 
allowable on the surface of a TTS may be determined by pressure equilibrium with
the surrounding non-magnetic (quiet) photosphere.  For Hubble 4, the 
equipartition
field strength is found to be $\sim 1.26$ kG in \S 3.2.2.  Thus, the mean
field of 2.51 kG found for this star is a factor of $\sim 2.0$ times too 
strong to be in pressure equipartition.  Model M4 has a 12\% filling factor
for 5 kG fields, which are a factor of $\sim 4$ too strong to be confined by
the gas pressure of the quiet photosphere.  However, pressure equilibrium
arguments (e.g. Spruit \& Zweibel 1979, Saar 1996, Safier 1999) only apply
if there are quiet regions on the star, and as discussed above, the data are
actually fit better if there are no such regions on the star.  Paper I
found that equipartition was violated on BP Tau and Saar (1996) found that
it is violated on a few very active K and M dwarfs.  While pressure 
equipartition has been suggested as a mechanism for determining the photospheric
magnetic field strength on cool stars (see also Rajaguru, Kurucz, \& Hasan
2002), it does not appear to hold for very active stars.  Indeed, it is 
expected that $B$ can exceed pressure equipartition when $f \approx 1$ 
(Solanki 1994).  In addition, the magnitude of velocity flows external to the
magnetic flux tubes may be important in determining both the distribution 
function (B\"unte \& Saar 1993) and magnitude (Solanki 1996) of magnetic 
field strengths in cool stars.  The solar chromosphere
and corona also violate equipartition, with the magnetic field dominating
the gas pressure in these regions.  Perhaps it should not be surprising that
very active stars extend this region of magnetic field dominance to deeper
layers of the atmosphere.  

The strong magnetic fields inferred on the 
surface of Hubble 4 dominate the gas pressure throughout the depth of the
photospheric model.  It is reasonable to ask if this magnetic dominance
continues to deep layers of the star, possibly altering its structure.
To investigate this, a polytropic stellar structure model was calculated
appropriate for fully convective stars (polytropic index 1.5, e.g. Clayton
1983).  As the calculation moves into the star, a decision must be made
regarding how to scale the magnetic field to account for the smaller 
surface area it fills at smaller radii.  Straight flux conservation would
imply scaling with $R^2$; however, there are no magnetic monopoles.  Dipolar
scaling could be used instead.  In either case, it turn out not to matter
significantly which is chosen as the gas pressure quickly become stronger
than the magnetic field pressure in the stellar structure calculation.
This occurs about $\sim 0.5$\% of the way into the star, or at a radius of
$R \sim 0.995R_*$.  So, while the photosphere of the star may be dominated
by magnetic pressure, the star as a whole is not.  In the photosphere though,
it is interesting to consider the appearance of star like Hubble 4 which
seems to be dominated by magnetic pressure.  Do the traditional solar 
features such as granulation, plage, and spots that are visible in the
solar photosphere also appear on such a magnetic star?  Spots clearly 
appear on these stars (e.g. Bouvier \& Bertout 1989), but it is almost certain
that these structures are modified by the strong fields.  Pevtsov et al. 
(2003) find that X-ray emission from TTSs is low as a group considering 
their strong magnetic fields.  This is confirmed by Feigelson et al. (2003)
who find that TTSs in the Orion Nebular Cluster saturate their X-ray emission
at a level an order of magnitude below where main sequence star do.  Perhaps
this is because convection is inhibited throughout the atmosphere of TTSs
due to their strong magnetic fields, in the same way convection is
diminished in sunspots.  The weaker convection may set a limit on the
mechanical energy available to heat the coronae of TTSs and produce X-rays.
If this is the case, studies of line bisectors in TTSs should show a systematic
difference relative to main sequence stars of the same spectral type;
however, such work is difficult due to line crowding at the typically
late spectral types of these stars and due to their relative faintness.

Safier (1999) originally called into question magnetic field
results for TTS based on the analysis of equivalent widths (e.g. Basri
et al. 1992, Guenther et al. 1999).  It is certainly true that great care
must be taken when using line equivalent widths to detect stellar
magnetic fields; however, the analysis presented above is based on
direct detection of Zeeman broadening in the spectral lines.  This
broadening shows that Hubble 4 has a large filling factor of very strong
($\sim 3$ kG based on the single component model) field.  On
the Sun, such strong field is found only in sunspots, which form deeper in
the photosphere than plage, due to lower gas density in the spot atmosphere.
It is therefore
interesting to consider whether the strong fields on TTS are confined
to large spots, particularly since large spots are routinely found in
photometric monitoring studies (e.g. Petrov et al. 1994) and in 
Doppler imaging studies (e.g. Johns--Krull \& Hatzes 1997) of TTS.
Guided by these studies, a model was constructed in which fields
stronger than could be supported by pressure equilibrium were assumed
to be in spots 1000 K cooler than the quiet photosphere.
The continuum difference between spotted and non-spotted regions acts
to increase the resulting filling factor of strong field regions (e.g.
Guenther et al. 1999); however, it was found that this effect was
essentially cancelled by the increase in the line strength of the
K band \ion{Ti}{1} lines formed in the cool spot atmosphere.  The spot
model field strength is an insignificant 8 G lower than that of the
single component model, and the filling factor is actually reduced by
13\%.  Such a spot model is successful at reproducing the IR observations
of the Ti lines; however, it predicts a very poor fit to the optical
spectrum, particularly in the TiO bandhead at 7055 \AA.  Thus it 
appears that the strong fields on Hubble 4 are not confined to cool
spots (or at least, not {\it only} confined to cool spots), but additional
work is needed to confirm this point.  

Realistic models of spot atmospheres are needed to make further progress on
testing whether cool spots contribute at all to the magnetic fields we have
detected on TTSs. The model used above
was simply that of a cooler star, with the same gravity and metallicity
as that found for Hubble 4.  On the Sun, magnetic pressure support
partially evacuates the gas in a spot, changing its density compared
to the quiet atmosphere.  Such effects need to be considered for
TTS (and other cool stars such as dMe stars).  In addition, since it
appears that the strong magnetic fields of TTS may also be present in
the quiet photosphere, atmospheric models taking into account the
magnetic pressure support should also be considered for this component of
the atmosphere.  Such a modeling effort is well beyond the scope of
the current paper, but these results should be regarded as an appeal to 
theoreticians to begin to consider such problems, particularly since the 
data presented here clearly indicates the presence of strong magnetic fields
on a large portion of Hubble 4.  Such strong fields have also been inferred
for a handful of other TTSs (Basri et al. 1992, Guenther et al. 1999, 
Paper I), and these strong fields are likely to be a common feature of TTSs
given their strong X-ray emission (e.g. Feigelson et al. 2003).

\acknowledgements

CMJ-K would like to acknowledge partial support from the NASA Origins
program through grant number NAG5-13103.  SHS would like to acknowledge
support from NSF grant 9528563 and a NASA Origins program grants
NAG5-10360 and NNG04GL54G.  We wish to thank the referee, Gibor Basri,
for his thoughtful comments on the original manuscript.

\clearpage



\newpage

\newpage

\begin{deluxetable}{lcccccc}
\tablewidth{14.5truecm}
\tablecaption{Observations\label{obs}}
\tablehead{
   \colhead{}&
   \colhead{Date}&
   \colhead{Start Time}&
   \colhead{}&
   \colhead{Total Exp.}&
   \colhead{}&
   \colhead{}\\[0.2ex]
   \colhead{Star}&
   \colhead{UT}&
   \colhead{UT}&
   \colhead{$N_{exp}$}&
   \colhead{Time (s)}&
   \colhead{Region}&
   \colhead{S/N}
}
\startdata
61 Cyg B & Dec. 12, 1996 & 5:19 & 6 & 300 & 2.2218 $\mu$m & 180 \\
HR 1044\tablenotemark{a} & Dec. 12, 1996 & 6:21 & 10 & 2000 & 2.2218 $\mu$m & 120 \\
Hubble 4 & Dec. 19, 1997 & 9:05 & 12 & 3600 & 2.3125 $\mu$m & 90 \\
HR 1666\tablenotemark{a} & Dec. 19, 1997 & 11:08 & 4 & 400 & 2.3125 $\mu$m & 250 \\
Hubble 4 & Dec. 19, 1997 & 13:07 & 10 & 3000 & 2.2218 $\mu$m & 60 \\
HR 1998\tablenotemark{a} & Dec. 19, 1997 & 14:07 & 4 & 960 & 2.2218 $\mu$m & 230 \\
Hubble 4 & Dec. 20, 1997 & 5:55 & 12 & 3600 & 2.2291 $\mu$m & 110 \\
HR 8781\tablenotemark{a} & Dec. 20, 1997 & 4:15 & 4 & 360 & 2.2291 $\mu$m & 250 \\
61 Cyg B & Dec. 5, 1998 & 6:17 & 2 & 240 & 2.2291 $\mu$m & 190 \\
Hubble 4 & Feb. 5, 1999 & 2:23 & 4 & 14400 & Optical & 60 \\
$\alpha$ Leo\tablenotemark{a} & Feb. 5, 1999 & 6:39 & 1 & 30 & Optical & 300 \\
\enddata
\tablenotetext{a}{Used as a telluric standard.}
\end{deluxetable}

\clearpage

\begin{deluxetable}{lrccc}
\tablewidth{11.0truecm}
\tablecaption{Best Fitting Nonmagnetic Model Parameters for Hubble 4\label{h4tab}}
\tablehead{
   \colhead{}&
   \colhead{\Teff}&
   \colhead{}&
   \colhead{}&
   \colhead{\vsini}\\[0.2ex]
   \colhead{}&
   \colhead{(K)}&
   \colhead{\logg}&
   \colhead{[M/H]}&
   \colhead{(\kms)}
}
\startdata
Fit to Data & 4158 & 3.61 & $-0.08$ & 14.6\hphantom{0}\\
Scatter from Data& 8 & 0.05 & $\hphantom{-}0.04$ & \hphantom{0}0.57\\
Uncert. from 61 Cyg B & 55 & 0.50 & $\hphantom{-}0.02$ & 1.6\\
Total Uncertainty & 56 & 0.50 & $\hphantom{-}0.04$ & \hphantom{0}1.7\hphantom{0}\\
\enddata
\end{deluxetable}

\begin{deluxetable}{lcc}
\tablewidth{7.0truecm}
\tablecaption{\protect{\ion{Ti}{1}} $gf$ Values\label{tigf}}
\tablehead{
   \colhead{$\lambda$(air)}&
   \colhead{Initial}&
   \colhead{61 Cyg B Fit}\\[0.2ex]
   \colhead{($\mu$m)}&
   \colhead{log($gf$)}&
   \colhead{log($gf$)}
}
\startdata
2.22112 & $-$1.770 & $-$1.637 \\
2.22328 & $-$1.658 & $-$1.558 \\
2.22740 & $-$1.756 & $-$1.632 \\
2.23106 & $-$2.124 & $-$1.940 \\
\enddata
\end{deluxetable}

\begin{deluxetable}{lcccccc}
\tablewidth{13.5truecm}
\tablecaption{Magnetic Fits\label{multi}}
\tablehead{
   \colhead{}&
   \colhead{S}&
   \colhead{M1}&
   \colhead{M2}&
   \colhead{M3}&
   \colhead{M4}&
   \colhead{SpP}
}
\startdata
$f_{0{\rm kG}}$& 0.29  & 0.01(01) & 0.14($<$01) & \nodata   & \nodata & 0.42 \\
$f_{1{\rm kG}}$&       & 0.34(03) & \nodata   & 0.36($<$01) & 0.37($<$01) & 0.00\\
$f_{2{\rm kG}}$&       & 0.02(04) & 0.52($<$01) & \nodata   & \nodata & \\
$f_{3{\rm kG}}$& 0.71  & 0.50(04) & \nodata   & 0.51($<$01) & 0.51($<$01) & 0.58 \\
$f_{4{\rm kG}}$&       & 0.06(02) & 0.34($<$01) & \nodata   & \nodata \\
$f_{5{\rm kG}}$&       & 0.01(01) & \nodata   & 0.12($<$01) & 0.12($<$01) \\
$f_{6{\rm kG}}$&       & 0.05(02) & 0.00($<$01) & \nodata   & \nodata \\
$f_{7{\rm kG}}$&       & 0.01(01) & \nodata   & 0.01($<$01) & \nodata \\
$\chi^2$       & 1.95 & 1.81 & 1.82 & 1.75 & 1.73 & 1.73 \\
$\Sigma B_i f_i$ (kG)& 2.12 & 2.56 & 2.40 & 2.53 & 2.51 & 1.72 \\
\enddata
\tablecomments{Model S is the single component model.  The actual field found
in the fit 2.98 kG.  Reported filling factors for the multi-component models
(M1--M4) are the mean of the 100 Monte Carlo trials for each model.  The 
$1\sigma$ uncertainty determined from the trials is given in parenthesis.  
The $\chi^2$ and $\Sigma B_i f_i$ values are for the specific distributions 
given above.  Model SpP is the spot plus plage model.  For the 1 kG 
component, the actual value use in the fit is the equipartition value of
1.26 kG.  The last component at 3 kG is the spot component (1,000 K cooler
than the quiet atmosphere), and the actual field value recovered from the
fit is 2.97 kG.}
\end{deluxetable}


\begin{references}
\reference{} Allard, F. \& Hauschildt, P.H. 1995, in Bottom of the Main
   Sequence and Beyond, ed. C.G. Tinney (Berlin: Springer), 32
\reference{} Basri, G., Marcy, G.\ W., \& Valenti, J.\ A. 1992,
   \apj, 390, 622
\reference{} Bevington, P.\ R. \& Robinson, D.\ K. 1992, Data Reduction and
   Error Analysis for the Physical Sciences, (New York: McGraw Hill)
\reference{} Bouvier, J. \& Bertout, C. 1989, \aap, 211, 99
\reference{} Brice{\~ n}o, C., Luhman, K.~L., Hartmann, L., Stauffer, J.~R.,
   \& Kirkpatrick, J.~D.\ 2002, \apj, 580, 317
\reference{} B\"unte, M., Saar, S.\ H.  1993, A\&A 271, 167
\reference{} Camenzind, M. 1990, Rev. Modern Astron.,
   (Berlin: Springer-Verlag), 3, 234
\reference{} Cameron, A.\ C. \& Campbell, C.\ G. 1993, \aap, 274, 309
\reference{} Clayton, D.\ D. 1983, Principles of Stellar Evolution and
   Nucleosynthesis, (Chicago: The University of Chicago Press)
\reference{} Edwards, S., Strom, S.\ E., Hartigan, P., Strom, K.\ M.,
   Hillenbrand, L.\ A., Herbst, W., Attridge, J., Merrill, K.\ M.,
   Probst, R., \& Gatley, I. 1993, \aj, 106, 372
\reference{} Feigelson, E.~D., Gaffney, J.~A., Garmire, G., Hillenbrand, 
   L.~A., \& Townsley, L.\ 2003, \apj, 584, 911 
\reference{} Goorvitch, D.\ 1994, \apjs, 95, 535
\reference{} Gray, D.\ F. 1992, The Observation and Analysis of Stellar 
   Photospheres, (Cambridge: Cambridge Univ. Press)
\reference{} Greene, T.~P., Tokunaga, A.~T., Toomey, D.~W., \& Carr, J.~B.\
   1993, \procspie, 1946, 313 
\reference{} Guenther, E.\ W., Lehmann, H., Emerson, J.\ P., \& Staude,
   1999, \aap, 341, 768
\reference{} Gullbring, E., Hartmann, L., Briceno, C., \& Calvet, N.\ 1998,
   \apj, 492, 323 
\reference{} Hartigan, P., Edwards, S., \& Ghandour, L.\ 1995, \apj, 452, 736 
\reference{} Hartmann, L., Hewett, R., \& Calvet, N. 1994, \apj, 426, 669
\reference{} Hartmann, L.~\& Stauffer, J.~R.\ 1989, \aj, 97, 873 
\reference{} Hatzes, A.\ P. 1995, \apj, 451, 784
\reference{} Herbst, W., Booth, J.\ F., Koret, D.\ L., Zajtseva, G.\ V., 
   Shakhovskaya, N.\ I., Vrba, F.\ J., Covino, E., Terranegra, L., 
   Vittone, A., Hoff, D., Kelsey, L., Lines, R., \& Barksdale, W. 1987, AJ,
   94, 137
\reference{} Herbst, W., Herbst, D.\ K., Grossman, E.\ J., \& Weinstein, D.
   1994, AJ, 108, 1906
\reference{} Hillenbrand, L.~A.~\& White, R.~J.\ 2004, \apj, 604, 741 
\reference{} Hinkle, K., Wallace, L., Valenti, J., \& Harmer, D.\ 2000, Visible
   and Near Infrared Atlas of the Arcturus Spectrum 3727-9300 \AA\ ed.~K.
   Hinkle, L. Wallace, J. Valenti, \& D. Harmer.~(San Francisco: ASP)
   ISBN: 1-58381-037-4
\reference{} Johns--Krull, C.\ M. \& Hatzes, A.\ P. 1997, \apj, 487, 896
\reference{} Johns--Krull, C.\ M. \& Valenti, J.\ A. 1996, \apjl, 459, L95
\reference{} Johns--Krull, C.~M.~\& Valenti, J.~A.\ 2001, \apj, 561, 1060 
\reference{} Johns--Krull, C.~M., Valenti, J.~A., Hatzes, A.~P., \& Kanaan, 
   A.\ 1999a, \apjl, 510, L41
\reference{} Johns--Krull, C.\ M., Valenti, J.\ A., \& Koresko, C. 1999b,
   \apj, 516, 900 (Paper I)
\reference{} Joncour, I., Bertout, C., \& Bouvier, J. 1994, \aap, 291, L19
\reference{} Joncour, I., Bertout, C., \& Menard, F. 1994, \aap, 285, L25
\reference{} K\"onigl, A. 1991, \apjl, 370, L39
\reference{} Leroy, J.\ L. 1962, Ann. d'Ap., 25, 127
\reference{} Livingston, W. \& Wallace, L. 1991, NSO Technical Report 
   \# 91-001
\reference{} McCarthy, J.~K., Sandiford, B.~A., Boyd, D., \& Booth, J.\ 1993,
   \pasp, 105, 881 
\reference{} M\'enard, F. \& Bertout, C. 1999, NATO ASIC Proc.~540: The 
   Origin of Stars and Planetary Systems, 341 
\reference{} Muzerolle, J., Calvet, N.; \& Hartmann, L. 1998, \apj, 492, 743
\reference{} Paatz, G.~\& Camenzind, M.\ 1996, \aap, 308, 77 
\reference{} Palla, F.~\& Stahler, S.~W.\ 1999, \apj, 525, 772
\reference{} Petrov, P.~P., Shcherbakov, V.~A., Berdyugina, S.~V., 
   Shevchenko, V.~S., Grankin, K.~N., \& Melnikov, S.~Y.\ 1994, \aaps, 107, 9 
\reference{} Pevtsov, A.~A., Fisher, G.~H., Acton, L.~W., Longcope, D.~W.,
   Johns--Krull, C.~M., Kankelborg, C.~C., \& Metcalf, T.~R. 2003, \apj, 598,
   1387
\reference{} Rajaguru, S.~P., Kurucz, R.~L., \& Hasan, S.~S.\ 2002, \apjl, 
   565, L101 
\reference{} Rice, J.\ B. \& Strassmeier, K.\ G. 1996, \aap, 316, 164
\reference{} Robinson, R.\ D. 1980, \apj, 239, 961
\reference{} R\"uedi, I., Solanki, S.~K., Livingston, W., \& Stenflo, J.~O.\ 
   1992, \aap, 263, 323
\reference{} Saar, S.\ H. 1988, \apj, 324, 441
\reference{} Saar, S.\ H. 1990, IAU Symp. 138, Solar Photosphere: Structure,
   Convection, and Magnetic Fields, ed.  J.\ O. Stenflo (Dordrecht: Kluwer),
   427
\reference{} Saar, S.\ H. 1994, IAU Symp. 154, Infrared Solar Physics, eds. 
   D.\ M. Rabin et al. (Dordrecht: Kluwer), 493
\reference{} Saar, S.\ H. 1996, IAU Symp. 176, Stellar Surface Structure, eds. 
   K.\ G. Strassmeier \& J.\ L. Linsky (Dordrecht: Kluwer), 237
\reference{} Saar, S.\ H. \& Linsky, J.\ L. 1985, \apj, 299, L47
\reference{} Safier, P.\ N. 1999, \apjl, 510, 127
\reference{} Shu, F.\ H., Najita, J., Ostriker, E., Wilkin, F., Ruden, S.,
   \& Lizano, S. 1994, \apj, 429, 781
\reference{} Skinner, S.~L.\ 1993, \apj, 408, 660
\reference{} Solanki, S.K. 1994, in Cool Stars, Stellar Systems, and the Sun,
   Eighth Cambridge Workshop. ASP Conference Series Vol. 64, ed. J.-P. 
   Caillault, 477
\reference{} Solanki, S.K. 1996, IAU Symp. 176, in Stellar Surface Structure,  
   eds. K.\ G. Strassmeier \& J.\ L.  (Dordrecht: Kluwer), 201
\reference{} Spruit, H.\ C., \& Zweibel, E.\ G. 1979, Sol. Phys., 62, 15
\reference{} Stassun, K.~G., Mathieu, R.~D., Mazeh, T., \& Vrba, F.~J.\ 1999,
   \aj, 117, 2941
\reference{} Strassmeier, K.\ G., Welty, A.\ D., \& Rice, J.\ B. 1994, \aap,
   285, L17
\reference{} Tokunaga, A.~T., Toomey, D.~W., Carr, J., Hall, D.~N.~B., \& 
   Epps, H.~W.\ 1990, \procspie, 1235, 131 
\reference{} Valenti, J.\ A., Marcy, G.\ W., \& Basri, G. 1995,
   \apj, 439, 939
\reference{} Valenti, J.~A., Piskunov, N., \& Johns-Krull, C.~M.\ 1998, \apj,
   498, 851

\end{references}
\end{document}